**Title:** Healthy Live Births Should be Considered as Competing Events when Estimating the Total Effect of Prenatal Medication Use on Pregnancy Outcomes

**Authors:** Chase D. Latour,[1,2*] Mark Klose,[1*] Jessie K. Edwards,[1] Zoey Song,[1] Michele Jonsson Funk,[1] Mollie E. Wood[1]
* These authors contributed equally to this work.

**Affiliations:**
[1] Department of Epidemiology, Gillings School of Global Public Health, University of North Carolina at Chapel Hill, Chapel Hill, NC, USA
[2] Cecil G. Sheps Center for Health Services Research, University of North Carolina at Chapel Hill, Chapel Hill, NC, USA

**Corresponding Author:**
Chase Latour
Department of Epidemiology
Gillings School of Global Public Health
University of North Carolina at Chapel Hill
135 Dauer Drive
2101 McGavran-Greenberg Hall
Chapel Hill, NC 27599-7435
chasedlatour@gmail.com

**Data Availability**: Data for this study were simulated using R statistical software (Vienna, Austria). All code is publicly available on GitHub: https://github.com/chasedlatour/Multiple_outcomes_pregnancy.

**Funding:** This research was partially supported by a National Research Service Award Pre-Doctoral/Post-Doctoral Traineeship from the Agency for Healthcare Research and Quality sponsored by The Cecil G. Sheps Center for Health Services Research, The University of North Carolina at Chapel Hill, Grant No. T32-HS000032 (CDL). This research was supported in part by a training grant from the National Institute of Child Health and Development [T32 HD052468] (CDL).

**Conflicts of Interest:** CDL has received payment from Target RWE and Regeneron Pharmaceuticals for unrelated work. Abbvie, Astellas, Boehringer Ingelheim, GlaxoSmithKline (GSK), Takeda, Sarepta, and UCB Bioscience have collaborative agreements with the Center for Pharmacoepidemiology housed in the Department of Epidemiology which provides salary support to MJF. MJF is a member of the Scientific Steering Committee (SSC) of a post-approval safety study of an unrelated drug class funded by GSK. All compensation for services provided on the SSC is invoiced by and paid to UNC Chapel Hill. MJF is a member of the Epidemiology and Clinical Advisory Board for Epividian. MEW is affiliated with the Center for Pharmacoepidemiology at the University of North Carolina at Chapel Hill and provides limited methods consulting to Center members unrelated to the present work. MK, JKE, and ZS have no other conflicts to disclose.

**Ethics Approval/Patient Consent:** Data for this study were simulated using statistical processes. No individuals' data were included in these analyses.

**ABSTRACT**

**Background:** Pregnancy loss is recognized as an important competing event in studies of prenatal medication use. However, a healthy live birth also precludes subsequent adverse pregnancy outcomes, yet these events are often censored.

**Objectives:** Using Monte Carlo simulation, examine bias that results from failure to account for healthy live birth as a competing event in estimates of the total effect of prenatal medication use on pregnancy outcomes.

**Methods:** We simulated 2,000 cohorts of 7,500 conceptions with chronic hypertension under 12 treatment effect profiles. Ongoing pregnancies were indexed into the trial and randomly assigned to initiate or not initiate antihypertensives. We used time-to-event methods to estimate absolute risks, risk differences (RDs) and risk ratios (RRs) for two outcomes: (1) composite fetal death or severe prenatal preeclampsia and (2) small-for-gestational-age (SGA) live birth. For the composite outcome, we conducted two analyses where non-preeclamptic live birth was (1) a censoring event and (2) a competing event. For SGA live birth, we conducted three analyses where fetal death and non-SGA live birth were (1) censoring events, (2) a competing event and censoring event, respectively; and (3) competing events.

**Results:** For the composite outcome, censoring non-preeclamptic live birth overestimated the absolute risk by 13 percentage points, on average, and produced biased RD and RR estimates (e.g., average bias in RD: 0.01). For SGA live birth, analyses censoring non-SGA live births (with or without fetal death was a competing event) overestimated absolute risk by 32 and 34 percentage points on average and generated biased RD and RR estimates (e.g., RD bias was 0.03 for both). Analyses in which healthy live births were modeled as competing events produced unbiased estimates of risks, RDs, and RRs.

**Conclusions:** Studies of prenatal exposures on pregnancy outcomes should analyze healthy live births as competing risks to estimate unbiased total treatment effects.

**KEY WORDS**

Pregnancy; pregnancy outcome; live birth; causality; prenatal exposure delayed effects; maternal exposure

**BACKGROUND**

Pregnancy is exceptionally sensitive to time, something that is particularly true when studying the effects of prenatal medication use on maternal and infant outcomes. For example, specific developmental events occur at predictable gestational ages.[1] Further, the cumulative incidence of pregnancy loss is substantial as pregnancies progress.[2,3] When a pregnancy ends at an earlier gestational age (e.g., 12 weeks via miscarriage), later outcomes (e.g., small for gestational age at delivery) are prevented.

Competing events, or events that prevent the occurrence of the study outcome, are common in pregnancy.[3–6] In studies of early pregnancy exposures on later outcomes, there is increasing recognition that miscarriages and induced abortions should be treated as competing events; failing to do so can induce substantial bias.[7–12] However, it is less often recognized that an event-free/healthy live birth can also function as a competing event.[13] For example, a live birth of an infant whose birthweight was in the 50th percentile at 37 gestational weeks cannot be delivered small for gestational age (SGA) at week 40.

The aim of this study was to evaluate the potential for bias in estimates of the total effect of treatment, in time-to-event analyses that treat healthy live births as censoring, rather than competing, events. We conducted a simulation study using a recently published clinical trial of antihypertensive treatment for mild-to-moderate chronic hypertension in pregnancy as a motivating example[14] and compared the effect of treatment on two outcomes – composite severe preeclampsia or fetal death and SGA at birth – using different nonparametric estimators of risk.

**METHODS**

*Simulation Study*

We simulated a simplified version of the Chronic Hypertension and Pregnancy (CHAP) clinical trial (Table 1).[14] Briefly, we simulated pregnancies with mild-to-moderate chronic hypertension that had not previously used antihypertensives. Simulated pregnancies were randomized to initiate or not initiate antihypertensives at gestational week 4 from conception (week 6 from LMP). We assumed perfect compliance and persistence of treatment and focused only on the intention-to-treat estimand.[15]

We generated two study outcomes based on the primary outcomes from the CHAP study. The first was a composite outcome of severe prenatal preeclampsia (i.e., preeclampsia with severe features that occurs prior to delivery) or fetal death at any gestational age. Pregnancies are at risk for this outcome from the start of follow-up through delivery, and a live delivery without severe preeclampsia prevents the patient from experiencing the outcome later (i.e., non-severe preeclamptic live birth is a competing event). The second outcome was delivery of a liveborn SGA infant. Both fetal death and live birth of a non-SGA infant prevent subsequent birth of a liveborn SGA infant. Patients are at risk for SGA live birth from gestational week 25 through delivery, and the competing events (fetal death or delivery to a non-SGA liveborn infant) prevent the pregnancy from experiencing the outcome later.

*Target Estimand*

Throughout this study, we focus on the total effect of treatment, defined as the effect of the treatment among those randomized at baseline, which aligns with the target estimand in the CHAP clinical trial. This is distinct from the controlled direct effect: the risk of an outcome under a scenario in which competing events do not occur or are prevented.[5,6,16,17] The total effect corresponds to the effect of an intervention that can be implemented in the real world and requires fewer identifying assumptions than the controlled direct effect.[17,18]

*Data Generation*

We conducted a Monte Carlo simulation study to investigate the bias induced by failing to account for healthy live births as competing events when estimating the total effect of medications in pregnancy. We generated data based on the directed acyclic graph in Figure 1 as described below (Figure S1).

*Step 1: Simulate pregnancies.*

We generated N=7,500 conceptions and simulated weekly potential pregnancy outcomes (fetal death, live birth, or continuing pregnancy) under "no treatment" for weeks 0 to 40 and "active treatment" for weeks 4 (trial enrollment) to 40. Weekly probabilities for potential outcomes under no antihypertensive use were selected to mirror published estimates.[2,14,19–24]

Potential outcomes under antihypertensive initiation incorporated the effect of initiation on the risk of miscarriage and preterm live birth. The weekly probabilities of fetal death from weeks 4 through 17 were determined via log-binomial regression:

$$p_{FD}(w) = \exp(\lambda_{0,w} + \lambda_1)$$

where $w$ is the week from conception, $p_{FD}(w)$ is the probability of fetal death at week $w$ (observed at week $w+1$), $\lambda_{0,w}$ is the log-probability of fetal death at $w$ among non-initiators, and $\lambda_1$ is the difference in the log-probability of fetal death among initiators versus non-initiators. The weekly probabilities of live birth were determined via log-binomial regression:

$$p_{LB}(w) = \begin{cases} 0 & w < 24 \\ \exp(\theta_{0,w} + \theta_1) & 24 \leq w < 34 \\ \exp(\theta_{0,w} + \theta_2) & w \geq 34 \end{cases}$$

where $w$ is the week from conception, $p_{LB}(w)$ the probability of live birth at week $w$ (observed at week $w+1$), $\theta_{0,w}$ the log-probability of live birth at $w$ among non-initiators, $\theta_1$ the difference in log-probability of live birth for initiators if $w < 34$, and $\theta_2$ the difference in log-probability of live birth for initiators if $w \geq 34$.

*Step 2: Determine treatment and subsequent outcomes.*

Some simulated pregnancies were not included in the trial because they experienced fetal death prior to trial entry (week 4) according to their potential outcomes under no antihypertensive use. Simulated pregnancies included in the trial were randomized in a 1:1 ratio to initiate or not initiate antihypertensive treatment.

*Step 3: Simulate development of severe preeclampsia.*

At gestational week 24 and beyond, simulated pregnancies could develop severe preeclampsia. We modeled the weekly probability of developing severe preeclampsia via logistic regression:

$$p_{PE}(w) = expit(\beta_{0,w} + \beta_1 a)$$

where $p_{PE}(w)$ is the probability of preeclampsia at week $w$, $\beta_{0,w}$ is the log-odds of preeclampsia at week $w$ among non-initiators, $\beta_1$ is the difference in log-odds of preeclampsia at week $w$ among initiators versus non-initiators, and $a$ is an indicator of being randomized to initiation ($a = 1$) or non-initiation ($a = 0$). For weeks 24 to 40, whether a patient developed preeclampsia was simulated by a Bernoulli random variable (i.e., $I_{PE,w} \sim Bern(p_{PE}(w))$ where $I_{PE,w}$ is an indicator for preeclampsia at week $w$ [observed at week $w + 1$]).

Simulated pregnancies that developed preeclampsia had outcomes regenerated such that they experienced a live birth or pregnancy loss within the following week. The probability depended on gestational age only (Supplemental File 2). We assumed preeclampsia occurred first.

*Step 4: Simulate small-for-gestational age values.*

SGA status was simulated among live births and assigned using a Bernoulli random variable (i.e., $I_{SGA} \sim Bern(p_{SGA})$ where $I_{SGA}$ is an indicator for an SGA infant and $p_{SGA}$ the probability of an SGA infant). $p_{SGA}$ was modeled using the below log-binomial regression:

$$p_{SGA} = \exp(\gamma_0 + \gamma_1 a + \gamma_2 I_{PE})$$

where $\gamma_0$ is the log-probability of a SGA infant among non-initiators without preeclampsia, $\gamma_1$ is the difference in the log-probability of a SGA infant among initiators versus non-initiators, $a$ is an indicator of treatment, $I_{PE}$ is an indicator for prenatal severe preeclampsia, and $\gamma_2$ is the difference in the log-probability of a SGA infant among pregnancies with preeclampsia compared to those without preeclampsia.

*Step 5: Randomly select pregnancies lost to follow-up.*

Because loss to follow-up is common in randomized trials and observational cohorts, simulated pregnancies had a 1% probability of being lost to follow-up (LTFU) at each gestational week after trial entry, necessitating time-to-event methods. Weekly probability of LTFU was independent of treatment and outcome (i.e., non-informative); this simplifying assumption isolates bias from censoring healthy live births.

*Step 6: Select observed outcomes.*

The above steps were used to generate all potential outcomes. Observed outcomes (pregnancy, severe prenatal preeclampsia, and SGA) depended upon the treatment assigned at randomization. Simulated pregnancies were determined to have experienced preeclampsia if the first preeclampsia event occurred the same week as or before their first pregnancy outcome. If preeclampsia occurred first, then their pregnancy outcome was determined based upon the preeclampsia-determined pregnancy outcomes. If the pregnancy outcome occurred first, then the simulated pregnancy did not develop severe prenatal preeclampsia and was assigned the corresponding pregnancy outcome. Data were then summarized into time-to-event and event status for both outcomes. If patients were LTFU, then their time-to-event was the censoring time, and they were marked as not having an outcome at that point.

We repeated this process 2,000 times for 12 scenarios distinguished by the effect of initiation on the risk of miscarriage, preterm birth, and SGA (Table 2). R code used to conduct these analyses is available on GitHub ([LINK TO FOLLOW UNBLINDING]).

### *Analyses Within Cohorts*

We conducted multiple analyses within each trial (Table S1).

*Kaplan-Meier Estimator*

The Kaplan-Meier estimator is a non-parametric estimator of cumulative incidence that can accommodate one outcome. It accounts for right censoring by imputing and redistributing censored individuals' eventual events to future event times according to the distribution of outcomes among uncensored individuals.[25,26] By censoring individuals at a competing event, the estimator assumes that those individuals who experienced a competing event eventually go on to experience the primary event with the same time-specific hazard as those remaining uncensored.[27,28] As a result, the Kaplan-Meier estimator consistently estimates the controlled direct effect of treatment on the study outcome when competing events are censored.

For the composite outcome, we treated fetal death or severe prenatal preeclampsia as an outcome event and all other event types (live births without prenatal severe preeclampsia, LTFU) as censoring events. For SGA live birth, deliveries of SGA infants were treated as an event and all other event types (fetal death, delivery to a non-SGA infant, LTFU) were censored. We calculated the cumulative risk in each treatment arm and estimated the treatment effect by contrasting these risks using the risk difference (RD) and risk ratio (RR).

*Aalen-Johansen Estimator*

The Aalen-Johansen estimator allows for multiple outcome types. Once someone experiences an outcome, that individual is removed from the risk set for other outcomes. As a result, the Aalen-Johansen estimator consistently estimates the total effect of an exposure on an outcome if all competing events are modeled appropriately.[29]

For the composite outcome (fetal death or prenatal severe preeclampsia), we considered live births without severe prenatal preeclampsia as competing events, and LTFU as a censoring event. For SGA live birth, we conducted two analyses: (1) SGA live birth as the outcome of interest, fetal death as a competing event, and both non-SGA live birth and LTFU as censoring events and (2) SGA live birth as the outcome of interest, fetal death and non-SGA live birth as separate competing events, and LTFU as a censoring event. These two analyses aimed to highlight the bias that may result if we only considered fetal death as a competing event. Like the Kaplan-Meier estimator, we contrasted the cumulative risks between the initiators versus non-initiators using the RD and RR.

*Kaplan-Meier Versus Aalen-Johansen Estimator*

Despite the KM estimator targeting a different causal estimand than the AJ estimator, we contrast these estimators because they would return the same absolute risk, RD, and RR estimates in the absence of competing events.[29] For example, in the case of the composite outcome, if the KM estimator (that censored healthy live births) returned different estimates than the AJ estimator (that treated healthy live births as competing events), it would imply that healthy live births were competing events.

*Recovering the Truth*

We compared our results to the effects calculated from the potential outcomes (i.e., the outcomes if everyone had initiated or not initiated) under full follow-up (i.e., no censoring). This corresponds to the true total effect of antihypertensive initiation versus non-initiation. These estimates were calculated within each trial; the truth was the average of these estimates across the 2,000 simulated trials.

*Summarizing Results Across Cohorts*

We summarized the parameter estimates from each of the 2,000 cohorts within a scenario by calculating the average estimate, bias, empirical standard error, and root mean squared error (Methods S3).[30]

**RESULTS**

The median number of pregnancies included in each clinical trial ranged from 5,482-5,484 across the 12 scenarios, which were evenly distributed across the two treatment arms (Table S2). For each scenario, between 17.1% and 20.0% of initiators and 18.5% and 18.6% of non-initiators were LTFU, with the increased variability among initiators due to the different effects of initiation on study outcomes.

*Composite Fetal Death and Prenatal Preeclampsia*

The Kaplan-Meier (KM) estimator overestimated the absolute risk of the composite outcome among initiators and non-initiators by 11-14 percentage points across scenarios (Figure 2, Tables S3 and S4). The Aalen-Johansen (AJ) estimator estimated the absolute risk of the composite outcome with little bias (range: 0 to 1 percentage point difference from the true absolute risk).

Inaccurate estimation of risks by the KM estimator resulted in biased contrasts of the risks. The RD and RR estimates from the KM estimator were biased in most scenarios, though the magnitude of bias tended to be small and towards the null (e.g., range of bias across scenarios in the RD: -0.02 to 0.01) (Figure 2, Table S3). As with the risks, RD and RR estimates from the AJ estimator were not biased. Scenario 2 (in which treatment increased the risk of fetal death and decreased the risk of preterm birth and SGA livebirth) demonstrated this pattern: the bias from the KM estimator was -0.02 for the RD and -0.06 for the (log-transformed) RR but 0.00 for both the RD and RR from the AJ estimator (Figure 2). The total effect estimates from the KM estimator that censored healthy live births diverged from those estimates from the AJ estimator that treated healthy live births as competing events.

The empirical standard error of the RD and RR estimates was much higher for the KM versus AJ estimator. For example, in scenario 2, the empirical standard errors were 0.09 and 0.01 for the RDs from the KM and AJ estimators, respectively.

*Small-for-Gestational-Age Live Birth*

As observed in analyses of the composite outcome, the KM estimator overestimated the risk of SGA livebirth for initiators and non-initiators, though by a larger margin (range: 30 to 36 percentage points difference) (Figure 3, Table S5). The AJ estimator that censored non-SGA live births similarly overestimated the absolute risks (range: 27 to 37 percentage points difference) while the AJ estimator with non-SGA live birth as a competing event did not (0 percentage points difference) (Tables S6-S7).

Both the KM estimator and AJ estimator that censored non-SGA live births produced biased RD and RR estimates, while the AJ estimator treating non-SGA live births as competing events did not (Figure 3). Considering again scenario 2, the bias in the RD and (log-transformed) RR estimates from the KM estimator were -0.04 and 0.14. The corresponding bias estimates for the AJ estimator that censored non-SGA live birth were -0.07 and 0.06. Only the AJ estimator treating non-SGA live birth as a competing event provided unbiased estimates. The magnitude and direction of bias in the KM estimator and AJ estimator censoring healthy live births was not consistent across scenarios and often depended upon the treatment effect estimate. The KM RD and RR estimates from scenario 2 provide an example: the RD is biased away from the null while the RR is biased towards the null. These differing directions derive from the inflated absolute risks: the same absolute difference does not represent the same relative difference if the baseline risk changes.

The AJ estimator treating non-SGA live birth as a competing event provided the least variable RD and RR estimates. For example, in scenario 2, the empirical standard error for the RD was 0.01. These corresponding values were 0.14 and 0.16 from the KM estimator and AJ estimator that censored non-SGA live births, respectively.

**COMMENT**

Our simulation study demonstrated that valid estimates of the absolute risk and total treatment effects could only be obtained from models that treated healthy live births as competing, rather than censoring, events. For both the composite and SGA livebirth outcomes, treating healthy live birth as a censoring event overestimated the risks. Similarly, RD and RR estimates from estimators that censored individuals at healthy live birth had more bias than those that modeled those live births as competing events. The direction and magnitude of bias varied according to the true treatment effect profile. Regardless of the bias, estimators that censored healthy live births provided less precise RD and RR estimates, illustrated by higher empirical standard errors.

Our results align with previous research on censoring competing events. Prior work has shown that censoring individuals at the occurrence of competing events results in higher absolute risk estimates than modeling those events as competing events, which was observed in our study.[4,13,25,31] This inflation in estimated risks occurs because time-to-event estimators assume that censored individuals go on to experience the study outcome. Prior simulation work has also found that censoring individuals at the time of a competing event will lead to less precise treatment effect estimates when those competing events are common, as we saw in our study.[32] Censoring the competing events reduces precision because the Kaplan-Meier estimator implicitly imputes outcomes for individuals who experienced a competing event based upon observed outcomes among a progressively smaller at-risk population. While it is fortunate that healthy live births are common, the high incidence of this specific competing event means that the choice of how to handle competing events is particularly consequential in studies of prenatal exposures.

The results of this study should be interpreted with the knowledge that statistical simulations are necessarily simplifications of real-world data. To avoid issues related to left truncation on the gestational age time scale, we indexed everyone at the same gestational age. This is impractical in real-world studies; investigators must incorporate statistical methodology that accounts for this left truncation.[33–37] Further, we simulated uninformative losses to follow-up; real studies, including trials, would need to consider statistical approaches to account for causes of censoring.[38] We also did not distinguish between induced versus spontaneous abortions: one CHAP trial inclusion criterion was the intention to not terminate pregnancy, so we assume the risk of termination is negligible. However, real studies must consider whether induced abortions represent a distinct competing event. Finally, all study outcomes were defined prior to or at delivery. Nonetheless, we emphasize that these issues persist for postpartum outcomes. For example, one CHAP study outcome was severe preeclampsia ≤2 weeks after delivery. Non-preeclamptic delivery would not be a competing event but surviving 2 weeks postpartum without the study outcome would: this would not change the conclusions from our analyses but merely delay the impacts by 2 weeks. Despite these limitations, the general results of our simulation will translate to more complicated settings, although differences in the magnitude and direction of the bias and degree of imprecision are expected. In addition to limitations in our data generating procedure, we only focused on the performance of non-parametric estimators for estimating risk but expect that our findings would extend to semi-parametric (e.g., Cox proportional hazards) or parametric (e.g., Weibull) models of time-to-event data.

There have been varied arguments around the treatment effect estimates that should be assessed in pregnancy medication studies.[17,18,39] This study focused on the estimation of total treatment effects. A strength of total effects is that the causal estimand aligns with a defined intervention and captures the effect of prenatal exposures on both pregnancy losses and delivery-related outcomes, among the population included at baseline. Some have argued that direct effects may be more relevant for clinical decision making.[39] However, there are significant challenges to identifying and interpreting these estimates.[17,18] First, the direct effect does not align with a clear intervention: it estimates the risk of an outcome if we were able to prevent all competing events without altering the hazard of the event of interest.[4–6,17] When miscarriage is the only competing event, imagining interventions to prevent miscarriage is appealing even if such interventions likely do not exist. However, when healthy live births are also a competing event, as was the case in our study, such interventions are not desirable. Second, obtaining a valid estimate of the direct effect requires controlling for causes of *both* the competing event and the study outcome, even in a trial. This is not required to estimate total effects, though investigators must still account for all confounders between treatment and the study outcome and competing event. Discussions about preferred estimands are outside the scope of this study but should be a focus of future work.

This study extends a sparse extant literature on using time to event methodology in pregnancy[35,37,40] by illustrating the importance of representing healthy live births as competing events when estimating total effects. Further, this work echoes prior calls to conceptualize pregnancy via multistate methodology. We considered healthy live births as one state of a multistate outcome.[13,41] For example, for SGA livebirth, observations could have one of three mutually exclusive outcomes: (1) fetal death, (2) livebirth of a non-SGA infant, and (3) live birth of an SGA infant. This work makes clear that investigators must assess whether healthy live births, or other positive pregnancy outcomes, function as competing events in their analyses.

**CONCLUSION**

We have shown that failure to account for healthy live birth as a competing event when estimating the total effects of prenatal medication exposures can lead to overestimated risks as well as biased and imprecise treatment effect estimates. The extent of this bias in real-world studies will depend on the magnitude of the underlying treatment effects and relationships between variables not explored here. Both absolute risks and treatment effects are necessary for informed decision-making between patients and providers, particularly during pregnancy when evidence from RCTs is often lacking. A doubling in risk may convey different health importance when the baseline risk is 1 in 1,000 versus 1 in 4 pregnancies. Studies of prenatal treatments should carefully consider the role of competing events, as well as the estimands implied by their analyses.

**TABLES AND FIGURE LEGENDS**

**Table 1.** Tabular description of the protocol components of the Chronic Hypertension and Pregnancy (CHAP) trial and our simulation.

| PROTOCOL COMPONENT | CHAP TRIAL | SIMULATED TRIAL |
|---|---|---|
| **ELIGIBILITY CRITERIA** | • Diagnosed with chronic hypertension in pregnancy<br>• Met blood pressure requirements at randomization<br>• Singleton pregnancy, without fetal anomaly, viable through 22 weeks of gestation<br>• Did not plan to terminate pregnancy<br>• No other exclusion criteria met | Same, limited to individuals who have not used antihypertensives prior to pregnancy. Further, individuals could only enter the trial at gestational week 4 from conception (6 from LMP). |
| **TREATMENT STRATEGIES** | Active treatment: Target blood pressure goal <140/90 mmHg, achieved by any therapy course deemed appropriate by the physician.<br><br>Standard treatment: Antihypertensive therapy withheld or stopped unless severe hypertension (≥160/105 mmHg) developed. | Active treatment: Initiate antihypertensive therapy.<br><br>Standard treatment: Do not initiate antihypertensive therapy. |
| **TREATMENT ASSIGNMENT** | Randomly assigned patients, stratified by center, to one of the two treatment strategies at screening. | Same, though pregnancies randomized at trial entry (gestational week 4 from conception, 6 from LMP). |
| **FOLLOW-UP PERIOD** | Follow-up started at random assignment to a treatment strategy. Patients followed from randomization through 6 weeks after their delivery. | Patients followed through the end of pregnancy. |
| **OUTCOMES** | Primary outcomes:<br>(1) Composite preeclampsia with severe features occurring up to 2 weeks after birth, medically indicated preterm birth before 35 weeks' gestation, placental abruption, or fetal or neonatal death. | Primary outcomes:<br>(1) Composite preeclampsia prior to delivery or fetal death.<br>(2) Small-for-gestational age live birth. |
|  | (2) Small-for-gestational age infant at delivery (<10$^{th}$ percentile). |  |
| **ESTIMAND** | Intention-to-treat effect | Same. |
| **ANALYSES** | Estimated risks of outcomes in each comparator group and compared them using the risk ratio and number needed to treat. Multiple imputation accounted for missing outcome values in primary analyses. A Kaplan-Meier estimator was used in supplemental analyses to construct the survival curve for time-from-randomization. Complete case analyses were conducted as a secondary approach. | Estimate risks of outcomes in each comparator group and comparing them using the risk difference and risk ratio. Time-to-event methodologies accounted for missing outcome values in primary analyses. |

**Figure 1.** Directed acyclic graph used for the data generation process. The variable $i$ indexes the gestational week. The dashed line indicates a deterministic outcome. Specifically, the dashed line between preeclampsia$_{i+1}$ and live birth$_{i+1}$ indicates that the occurrence of preeclampsia induced a pregnancy outcome. The dashed line between live birth$_{i+1}$ and SGA$_{i+1}$ indicates that small-for-gestational age was only generated among live births.

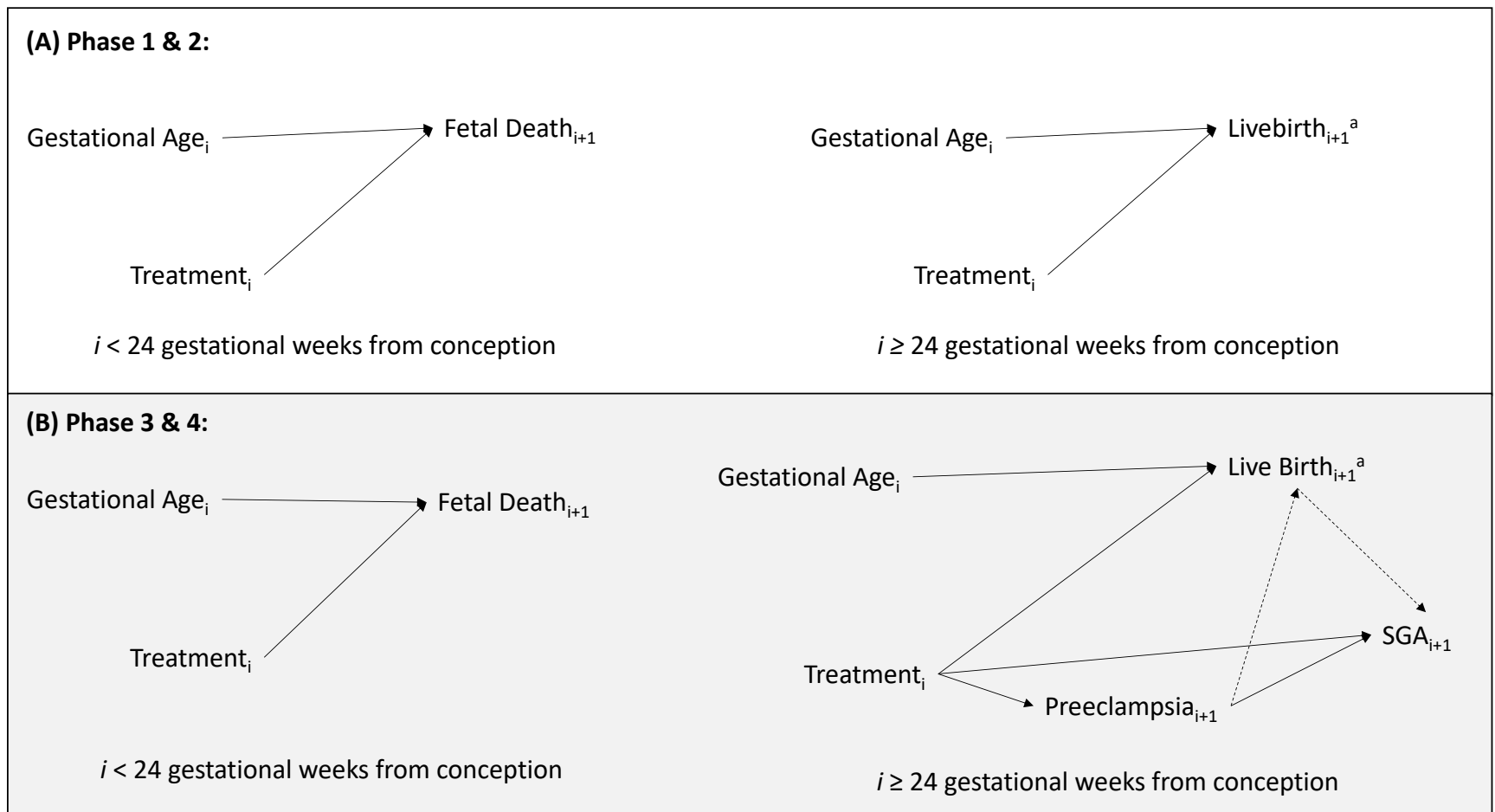

[a] The other option is fetal death, but only one is shown, as live birth is only possible in these gestational weeks.

**Table 2**. Tabular description of all the scenarios considered in the analyses. Notation regarding parameter values corresponds to the equations provided in the Methods.

| | | Effect of Treatment on… | | |
|---|---|---|---|---|
| **Scenario** | **Miscarriage** | **Preterm Birth** | **Severe prenatal preeclampsia**[c] | **Small-for-gestational age**[d] |
| **Scenario 1** | Decreases risk of miscarriage ($\lambda = \log(0.8)$) | Decreases risk of preterm birth ($\theta_1 = \log(0.7)$, $\theta_2 = \log(1.05)$) | Decreases risk of severe preeclampsia ($\beta_1 = \log(0.6)$) | Decreases risk of small-for-gestational age ($\lambda_1 = \log(0.8)$) |
| **Scenario 2** | Increases risk of miscarriage ($\lambda = \log(1.25)$) | | | |
| **Scenario 3** | No impact on risk of miscarriage ($\lambda = 0$) | | | |
| **Scenario 4** | Decreases risk of miscarriage ($\lambda = \log(0.8)$) | No effect on risk of preterm birth ($\theta_1 = \theta_2 = 0$) | | |
| **Scenario 5** | Increases risk of miscarriage ($\lambda = \log(1.25)$) | | | |
| **Scenario 6** | No impact on risk of miscarriage ($\lambda = 0$) | | | |
| **Scenario 7** | Decreases risk of miscarriage ($\lambda = \log(0.8)$) | Decreases risk of preterm birth ($\theta_1 = \log(0.7)$, $\theta_2 = \log(1.05)$) | | No effect on risk of small-for-gestational age |
| **Scenario 8** | Increases risk of miscarriage ($\lambda = \log(1.25)$) | | | |
| **Scenario 9** | No impact on risk of miscarriage ($\lambda = 0$) | | | |
| **Scenario 10** | Decreases risk of miscarriage ($\lambda = \log(0.8)$) | No effect on risk of preterm birth ($\theta_1 = \theta_2 = 0$) | | |
| **Scenario 11** | Increases risk of miscarriage ($\lambda = \log(1.25)$) | | | |
| **Scenario 12** | No impact on risk of miscarriage ($\lambda = 0$) | | | |

RR = risk ratio.

Note: Notation corresponds to formulae provided in the Methods section.

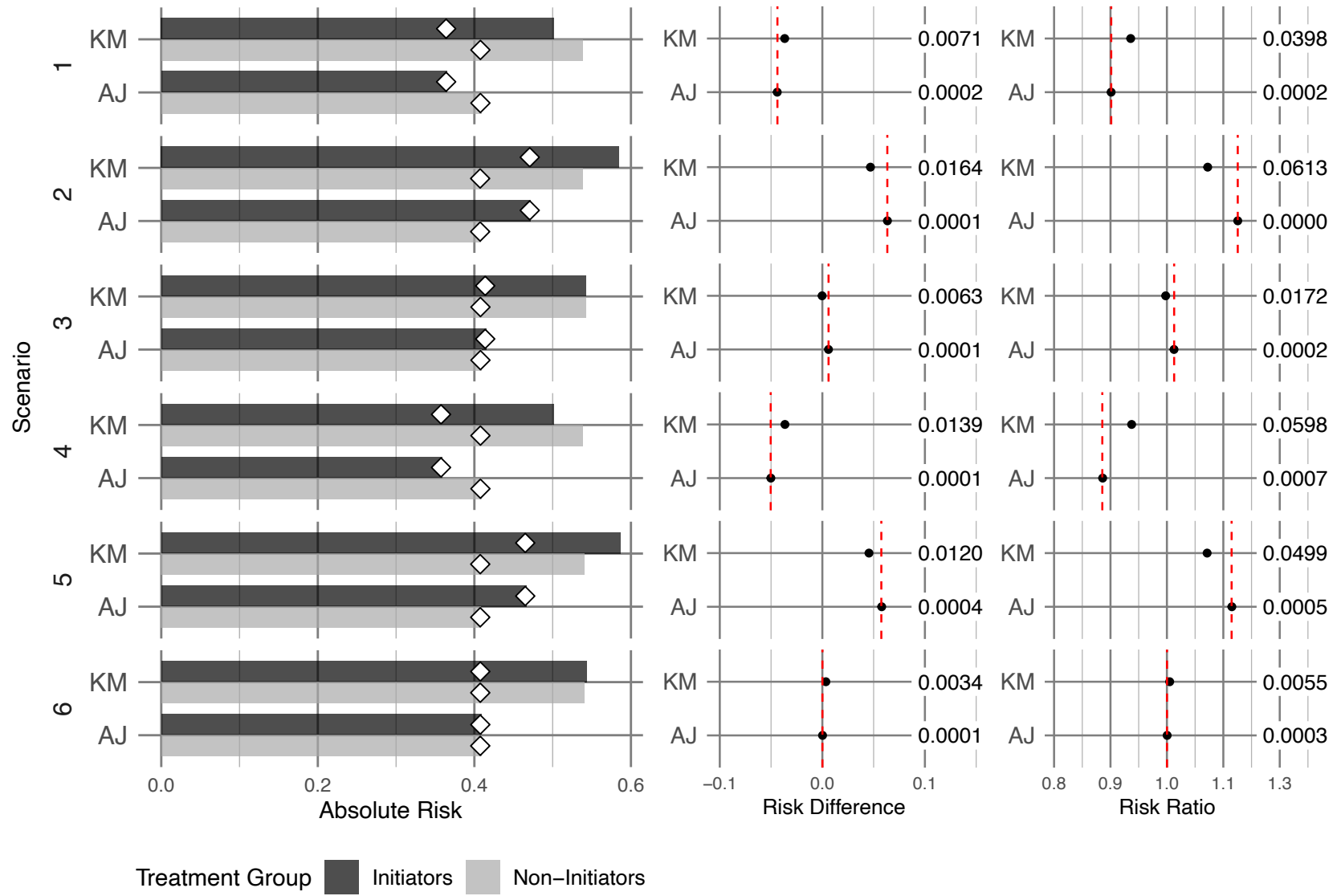

**Figure 2.** Risk, risk difference, and risk ratio estimates from the Kaplan-Meier (KM) and Aalen-Johansen (AJ) estimators for the composite outcome of fetal death or severe prenatal preeclampsia for Scenarios 1-6. Results were identical for Scenarios 7-12. The bars on the left-hand side represent the observed absolute risks, and the overlaid diamonds represent the true absolute risks. The dashed line next to the risk difference and risk ratio estimates represents the truth, according to the potential outcomes. The numbers to the right of the risk difference and risk ratio estimates correspond to their absolute bias from the truth. Scenarios are differentiated according to the effect of initiation on the risk of miscarriage and preterm live birth. Initiation decreases the risk of preterm birth across scenarios 1-3, while it decreases the risk of miscarriage in scenario 1, increases the risk in scenario 2, and does not affect the risk in scenario 3. Initiation has no effect on the risk of preterm birth across scenarios 4-6 with the same patterning of the effects on the risk of miscarriage.

**Figure 3.** Risk, risk difference, and risk ratio estimates from the Kaplan-Meier (KM) and Aalen-Johansen (AJ) estimators for small-for-gestational-age (SGA) for Scenarios 1-12. The bars on the left-hand side represent the observed absolute risks, and the overlaid diamonds represent the true absolute risks. The dashed line next to the risk difference and risk ratio estimates represents the truth, according to the potential outcomes. The numbers to the right of the risk difference and risk ratio estimates correspond to their absolute bias from the truth. Scenarios are differentiated according to the effect of initiation on the risk of miscarriage, preterm live birth, and SGA with initiation decreasing the risk of SGA in scenarios 1-6 and having no effect on the risk of SGA in scenarios 7-12. Initiation decreases the risk of preterm birth for scenarios 1-3 and 7-9 and has no effect on the risk of preterm live birth in scenarios 4-6 and 10-12. Initiation decreases the risk of miscarriage in scenarios 1, 4, 7, and 10; increases the risk in scenarios 2, 5, 8, and 11; and has no effect on the risk in scenarios 3, 6, 9, and 12.

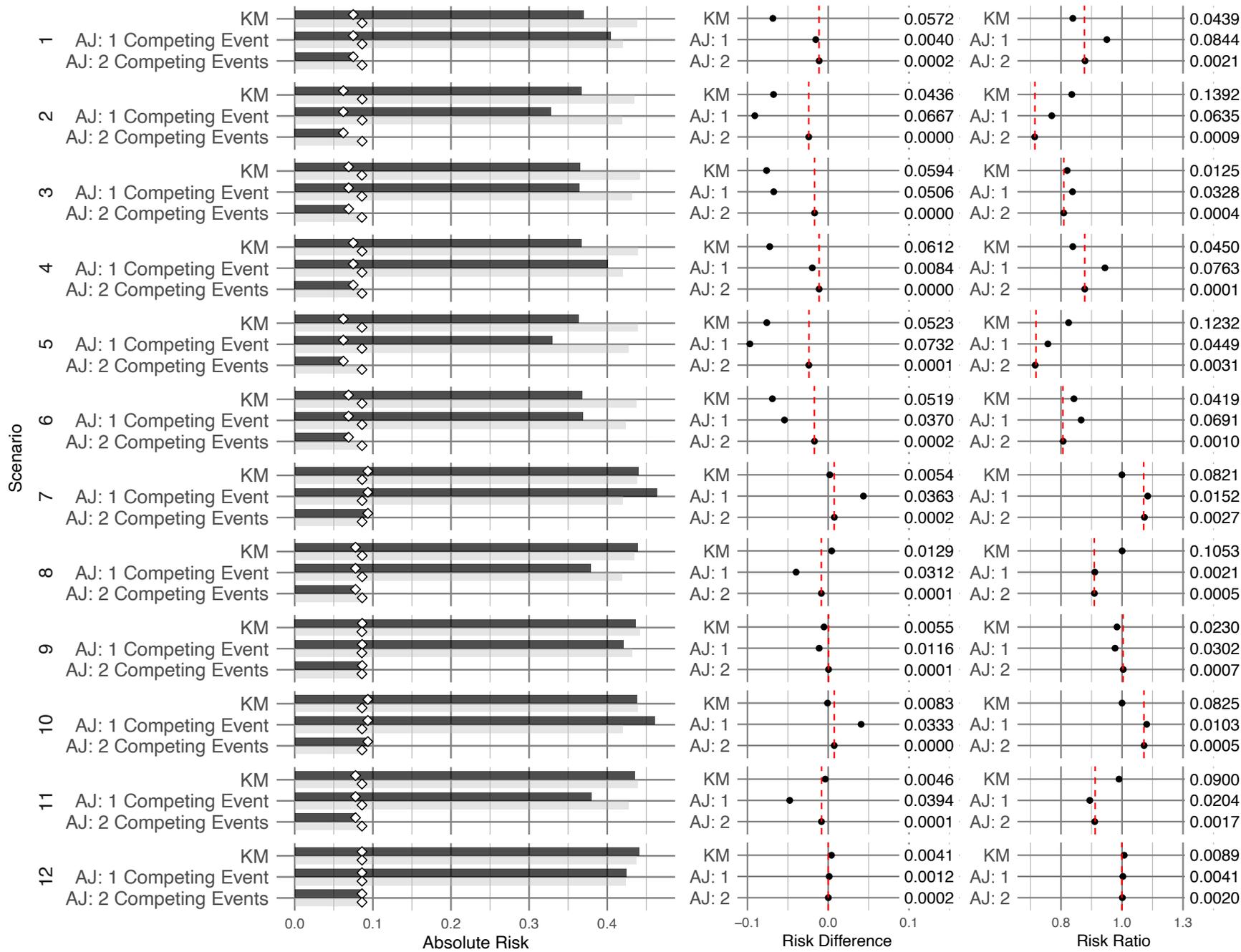

**SUPPLEMENTAL MATERIAL**

**Figures**

*Figure S1. Depiction of the data generation process for an example patient. This patient was randomly assigned to initiate antihypertensives. Their observed outcomes then would be that they developed severe preeclampsia, observed at week 37 from conception, but ultimately experienced a live birth to a small-for-gestational-age infant. However, that patient would have been censored at gestational week 30 from conception, making their outcome unobserved. If the patient had been assigned to not initiate treatment, they would have experienced fetal death, observed at gestational week 12 from conception.*

***Step 1:*** *Simulate a cohort of pregnant people that could be included in the trial.*

| Weeks from Conception | 0 | 1 | 2 | 3 | 4 | 5 | 6 | 7 | 8 | 9 | 10 | 11 | 12 | 13 | 14 | 15 | 16 | 17 | 18 | 19 | 20 | 21 | 22 | 23 | 24 | 25 | 26 | 27 | 28 | 29 | 30 | 31 | 32 | 33 | 34 | 35 | 36 | 37 | 38 | 39 | 40 |
|---|---|---|---|---|---|---|---|---|---|---|---|---|---|---|---|---|---|---|---|---|---|---|---|---|---|---|---|---|---|---|---|---|---|---|---|---|---|---|---|---|---|
| Pregnancy Outcomes Under No Initiation | C | C | C | C | C | C | C | C | C | C | C | C | FD | C | C | C | FD | C | C | C | C | C | C | C | C | C | C | C | C | C | C | C | C | C | C | C | C | LB | LB | C | C |
| Pregnancy Outcomes Under Initiation | C | C | C | C | C | C | C | C | C | C | C | C | C | C | C | C | C | C | C | C | C | C | C | C | C | C | C | C | C | C | C | C | C | C | C | C | C | C | LB | C | C |

***Step 2:*** *Determine treatment and subsequent outcomes based upon treatment.*

Treatment Arm: Initiate Antihypertensives

| Weeks from Conception | 0 | 1 | 2 | 3 | 4 | 5 | 6 | 7 | 8 | 9 | 10 | 11 | 12 | 13 | 14 | 15 | 16 | 17 | 18 | 19 | 20 | 21 | 22 | 23 | 24 | 25 | 26 | 27 | 28 | 29 | 30 | 31 | 32 | 33 | 34 | 35 | 36 | 37 | 38 | 39 | 40 |
|---|---|---|---|---|---|---|---|---|---|---|---|---|---|---|---|---|---|---|---|---|---|---|---|---|---|---|---|---|---|---|---|---|---|---|---|---|---|---|---|---|---|
| Observed Pregnancy Outcomes Under Initiation | C | C | C | C | C | C | C | C | C | C | C | C | C | C | C | C | C | C | C | C | C | C | C | C | C | C | C | C | C | C | C | C | C | C | C | C | C | C | LB | C | C |

***Step 3:*** *Simulate development of preeclampsia. This was done for both potential outcomes.*

| | 0 | 1 | 2 | 3 | 4 | 5 | 6 | 7 | 8 | 9 | 10 | 11 | 12 | 13 | 14 | 15 | 16 | 17 | 18 | 19 | 20 | 21 | 22 | 23 | 24 | 25 | 26 | 27 | 28 | 29 | 30 | 31 | 32 | 33 | 34 | 35 | 36 | 37 | 38 | 39 | 40 |
|---|---|---|---|---|---|---|---|---|---|---|---|---|---|---|---|---|---|---|---|---|---|---|---|---|---|---|---|---|---|---|---|---|---|---|---|---|---|---|---|---|---|
| Preeclampsia \| No Initiation | 0 | 0 | 0 | 0 | 0 | 0 | 0 | 0 | 0 | 0 | 0 | 0 | 0 | 0 | 0 | 0 | 0 | 0 | 0 | 0 | 0 | 0 | 0 | 0 | 0 | 0 | 0 | 0 | 0 | 0 | 1 | 0 | 0 | 0 | 0 | 0 | 0 | 1 | 0 | 0 | 0 |
| Preeclampsia \| Initiation | 0 | 0 | 0 | 0 | 0 | 0 | 0 | 0 | 0 | 0 | 0 | 0 | 0 | 0 | 0 | 0 | 0 | 0 | 0 | 0 | 0 | 0 | 0 | 0 | 0 | 0 | 0 | 0 | 0 | 0 | 0 | 0 | 0 | 0 | 0 | 0 | 0 | 1 | 0 | 0 | 0 |
| Pregnancy Outcome after Preeclampsia | | | | | | | | | | | | | | | | | | | | | | | | | | | FD | FD | FD | LB | FD | LB | LB | FD | LB | LB | LB | LB | LB | LB | LB |

***Step 4:*** *Simulate patients' small-for-gestational age values.*

SGA | No Initiation = 0 – Fetal death if untreated, so not generated.
SGA | Initiation = 1 – Randomly generated since a live birth if treated.

***Step 5:*** *Randomly select patients to be censored.*

| | 0 | 1 | 2 | 3 | 4 | 5 | 6 | 7 | 8 | 9 | 10 | 11 | 12 | 13 | 14 | 15 | 16 | 17 | 18 | 19 | 20 | 21 | 22 | 23 | 24 | 25 | 26 | 27 | 28 | 29 | 30 | 31 | 32 | 33 | 34 | 35 | 36 | 37 | 38 | 39 | 40 |
|---|---|---|---|---|---|---|---|---|---|---|---|---|---|---|---|---|---|---|---|---|---|---|---|---|---|---|---|---|---|---|---|---|---|---|---|---|---|---|---|---|---|
| Indicator of censoring | 0 | 0 | 0 | 0 | 0 | 0 | 0 | 0 | 0 | 0 | 0 | 0 | 0 | 0 | 0 | 0 | 0 | 0 | 0 | 0 | 0 | 0 | 0 | 0 | 0 | 0 | 0 | 0 | 0 | 0 | 1 | 0 | 0 | 0 | 0 | 0 | 0 | 0 | 0 | 1 | 1 |

C = Continuing pregnancy to the next gestational week; FD = fetal death (observed at the next gestational week); LB = live birth (observed at the next gestational week)

**Tables**

*Table S1. Tabular description of all the analyses conducted within each cohort.*

| Study Outcome | Estimator | Competing Event | Censoring Event |
|---|---|---|---|
| Composite fetal death and severe prenatal preeclampsia | Kaplan- Meier | NA | Non-preeclamptic live birth, loss to follow-up (LTFU) |
| | Aalen-Johansen | Non-preeclamptic live birth | LTFU |
| SGA Infant | Kaplan-Meier | NA | Fetal death, Non-SGA live birth, LTFU |
| | Aalen-Johansen (1) | Fetal Death | Non-SGA live birth, LTFU |
| | Aalen-Johansen (2) | Fetal Death, Non-SGA live birth | LTFU |

Table S2. Descriptive statistics of the 2,000 cohorts generated for each data generating scenario. Counts represent the median number of pregnancies that met that description across the 2,000 cohorts for a set of data generating parameters. The percentages represent the distribution across initiators versus non-initiators.

| Scenario | Number of Pregnancies Included in the Trial | | Number of Censored Pregnancies | | Number of Fetal Deaths, Ignoring Censoring | | Number of Live Births, Ignoring Censoring | |
|---|---|---|---|---|---|---|---|---|
| | Non-Initiators | Initiators | Non-Initiators | Initiators | Non-Initiators | Initiators | Non-Initiators | Initiators |
| 1 | 2742 (50%) | 2742 (50%) | 509 (48%) | 549 (52%) | 905 (55%) | 751 (45%) | 1836 (48%) | 1991 (52%) |
| 2 | 2742 (50%) | 2742 (50%) | 509 (52%) | 476 (48%) | 904 (45%) | 1084 (55%) | 1837 (53%) | 1657 (47%) |
| 3 | 2741 (50%) | 2741 (50%) | 509 (50%) | 516 (50%) | 905 (50%) | 906 (50%) | 1837 (50%) | 1833 (50%) |
| 4 | 2741 (50%) | 2741 (50%) | 509 (48%) | 542 (52%) | 906.5 (55%) | 751 (45%) | 1836.5 (48%) | 1991 (52%) |
| 5 | 2741 (50%) | 2742 (50%) | 509 (52%) | 469 (48%) | 904 (45%) | 1083 (55%) | 1837 (53%) | 1659 (47%) |
| 6 | 2742 (50%) | 2742 (50%) | 508 (50%) | 509 (50%) | 906 (50%) | 905 (50%) | 1836 (50%) | 1836 (50%) |
| 7 | 2742 (50%) | 2742 (50%) | 509 (48%) | 549 (52%) | 905 (55%) | 751 (45%) | 1836 (48%) | 1991 (52%) |
| 8 | 2742 (50%) | 2742 (50%) | 509 (52%) | 476 (48%) | 904 (45%) | 1084 (55%) | 1837 (53%) | 1657 (47%) |
| 9 | 2741 (50%) | 2741 (50%) | 509 (50%) | 516 (50%) | 905 (50%) | 906 (50%) | 1837 (50%) | 1833 (50%) |
| 10 | 2741 (50%) | 2741 (50%) | 509 (48%) | 542 (52%) | 906.5 (55%) | 751 (45%) | 1836.5 (48%) | 1991 (52%) |
| 11 | 2741 (50%) | 2742 (50%) | 509 (52%) | 469 (48%) | 904 (45%) | 1083 (55%) | 1837 (53%) | 1659 (47%) |
| 12 | 2742 (50%) | 2742 (50%) | 508 (50%) | 509 (50%) | 906 (50%) | 905 (50%) | 1836 (50%) | 1836 (50%) |

*Table S3. Results of the analyses for the composite outcome of fetal death or severe prenatal preeclampsia across all 12 scenarios estimated via a Kaplan-Meier estimator where live birth without severe prenatal preeclampsia was treated as a censoring event.*

| | Effect of Treatment on Risk of… | | | Truth | | | | Kaplan-Meier Estimator | | | | | | | | |
|---|---|---|---|---|---|---|---|---|---|---|---|---|---|---|---|---|
| | | | | Risk | | | | Risk | | Risk Difference | | | | Risk Ratio[a] | | | |
| Scenario | Miscarriage | Preterm live birth | SGA | Initiators | Non-Initiators | RD | RR | Initiators | Non-Initiators | Estimate | Bias | ESE | RMSE | Estimate | Bias | ESE | RMSE |
| 1 | ↓ | ↓ | ↓ | 0.36 | 0.41 | -0.04 | 0.89 | 0.50 | 0.54 | -0.04 | 0.01 | 0.09 | 0.09 | 0.93 | 0.04 | 0.16 | 0.17 |
| 2 | ↑ | ↓ | ↓ | 0.47 | 0.41 | 0.06 | 1.16 | 0.58 | 0.54 | 0.05 | -0.02 | 0.09 | 0.09 | 1.09 | -0.06 | 0.14 | 0.16 |
| 3 | --- | ↓ | ↓ | 0.41 | 0.41 | 0.01 | 1.01 | 0.54 | 0.54 | 0.00 | -0.01 | 0.09 | 0.09 | 1.00 | -0.02 | 0.16 | 0.16 |
| 4 | ↓ | --- | ↓ | 0.36 | 0.41 | -0.05 | 0.88 | 0.50 | 0.54 | -0.04 | 0.01 | 0.08 | 0.09 | 0.93 | 0.06 | 0.16 | 0.17 |
| 5 | ↑ | --- | ↓ | 0.46 | 0.41 | 0.06 | 1.14 | 0.59 | 0.54 | 0.05 | -0.01 | 0.08 | 0.08 | 1.09 | -0.05 | 0.14 | 0.15 |
| 6 | --- | --- | ↓ | 0.41 | 0.41 | 0.00 | 1.00 | 0.54 | 0.54 | 0.00 | 0.00 | 0.08 | 0.08 | 1.01 | 0.01 | 0.14 | 0.14 |
| 7 | ↓ | ↓ | --- | 0.36 | 0.41 | -0.04 | 0.89 | 0.50 | 0.54 | -0.04 | 0.01 | 0.09 | 0.09 | 0.93 | 0.04 | 0.16 | 0.17 |
| 8 | ↑ | ↓ | --- | 0.47 | 0.41 | 0.06 | 1.16 | 0.58 | 0.54 | 0.05 | -0.02 | 0.09 | 0.09 | 1.09 | -0.06 | 0.14 | 0.16 |
| 9 | --- | ↓ | --- | 0.41 | 0.41 | 0.01 | 1.01 | 0.54 | 0.54 | 0.00 | -0.01 | 0.09 | 0.09 | 1.00 | -0.02 | 0.16 | 0.16 |
| 10 | ↓ | --- | --- | 0.36 | 0.41 | -0.05 | 0.88 | 0.50 | 0.54 | -0.04 | 0.01 | 0.08 | 0.09 | 0.93 | 0.06 | 0.16 | 0.17 |
| 11 | ↑ | --- | --- | 0.46 | 0.41 | 0.06 | 1.14 | 0.59 | 0.54 | 0.05 | -0.01 | 0.08 | 0.08 | 1.09 | -0.05 | 0.14 | 0.15 |
| 12 | --- | --- | --- | 0.41 | 0.41 | 0.00 | 1.00 | 0.54 | 0.54 | 0.00 | 0.00 | 0.08 | 0.08 | 1.01 | 0.01 | 0.14 | 0.14 |

SGA = small-for-gestational age; RD = risk difference; RR = risk ratio, ESE = empirical standard error, RMSE = root mean squared error

[a] Bias, empirical standard error, and root mean squared error were calculated using the natural log-transformed risk ratio and reported on that scale.

*Table S4. Results of the analyses for the composite outcome of fetal death or severe prenatal preeclampsia across all 12 scenarios estimated via an Aalen-Johansen estimator where live birth without severe preeclampsia was modeled as a competing event.*

| | Effect of Treatment on Risk of… | | | Truth | | | | Aalen-Johansen Estimator | | | | | | Risk Ratio[a] | | | |
|---|---|---|---|---|---|---|---|---|---|---|---|---|---|---|---|---|---|
| | | | | Risk | | | | Risk | | Risk Difference | | | | | | | |
| Scenario | Miscarriage | Preterm live birth | SGA | Initiators | Non-Initiators | RD | RR | Initiators | Non-Initiators | Estimate | Bias | ESE | RMSE | Estimate | Bias | ESE | RMSE |
| 1 | ↓ | ↓ | ↓ | 0.36 | 0.41 | -0.04 | 0.89 | 0.36 | 0.41 | -0.04 | 0.00 | 0.01 | 0.01 | 0.89 | 0.00 | 0.04 | 0.04 |
| 2 | ↑ | ↓ | ↓ | 0.47 | 0.41 | 0.06 | 1.16 | 0.47 | 0.41 | 0.06 | 0.00 | 0.01 | 0.01 | 1.16 | 0.00 | 0.03 | 0.03 |
| 3 | --- | ↓ | ↓ | 0.41 | 0.41 | 0.01 | 1.01 | 0.42 | 0.41 | 0.01 | 0.00 | 0.01 | 0.01 | 1.01 | 0.00 | 0.03 | 0.03 |
| 4 | ↓ | --- | ↓ | 0.36 | 0.41 | -0.05 | 0.88 | 0.36 | 0.41 | -0.05 | 0.00 | 0.01 | 0.01 | 0.88 | 0.00 | 0.04 | 0.04 |
| 5 | ↑ | --- | ↓ | 0.46 | 0.41 | 0.06 | 1.14 | 0.47 | 0.41 | 0.06 | 0.00 | 0.01 | 0.01 | 1.14 | 0.00 | 0.03 | 0.03 |
| 6 | --- | --- | ↓ | 0.41 | 0.41 | 0.00 | 1.00 | 0.41 | 0.41 | 0.00 | 0.00 | 0.01 | 0.01 | 1.00 | 0.00 | 0.03 | 0.03 |
| 7 | ↓ | ↓ | --- | 0.36 | 0.41 | -0.04 | 0.89 | 0.36 | 0.41 | -0.04 | 0.00 | 0.01 | 0.01 | 0.89 | 0.00 | 0.04 | 0.04 |
| 8 | ↑ | ↓ | --- | 0.47 | 0.41 | 0.06 | 1.16 | 0.47 | 0.41 | 0.06 | 0.00 | 0.01 | 0.01 | 1.16 | 0.00 | 0.03 | 0.03 |
| 9 | --- | ↓ | --- | 0.41 | 0.41 | 0.01 | 1.01 | 0.42 | 0.41 | 0.01 | 0.00 | 0.01 | 0.01 | 1.01 | 0.00 | 0.03 | 0.03 |
| 10 | ↓ | --- | --- | 0.36 | 0.41 | -0.05 | 0.88 | 0.36 | 0.41 | -0.05 | 0.00 | 0.01 | 0.01 | 0.88 | 0.00 | 0.04 | 0.04 |
| 11 | ↑ | --- | --- | 0.46 | 0.41 | 0.06 | 1.14 | 0.47 | 0.41 | 0.06 | 0.00 | 0.01 | 0.01 | 1.14 | 0.00 | 0.03 | 0.03 |
| 12 | --- | --- | --- | 0.41 | 0.41 | 0.00 | 1.00 | 0.41 | 0.41 | 0.00 | 0.00 | 0.01 | 0.01 | 1.00 | 0.00 | 0.03 | 0.03 |

SGA = small-for-gestational age; RD = risk difference; RR = risk ratio, ESE = empirical standard error, RMSE = root mean squared error

[a] Bias, empirical standard error, and root mean squared error were calculated using the natural log-transformed risk ratio and reported on that scale.

*Table S5. Results of the analyses for the small-for-gestational-age (SGA) live birth across all 12 scenarios estimated via a Kaplan-Meier estimator.*

|  | | | | Truth | | | | Kaplan-Meier Estimator | | | | | | | | |
|---|---|---|---|---|---|---|---|---|---|---|---|---|---|---|---|---|
|  | Effect of Treatment on Risk of… | | | Risk | | | | Risk | | Risk Difference | | | | Risk Ratio[a] | | | |
| Scenario | Miscarriage | Preterm live birth | SGA | Initiators | Non-Initiators | RD | RR | Initiators | Non-Initiators | Estimate | Bias | ESE | RMSE | Estimate | Bias | ESE | RMSE |
| 1 | ↓ | ↓ | ↓ | 0.07 | 0.09 | -0.01 | 0.87 | 0.37 | 0.44 | -0.07 | -0.06 | 0.14 | 0.15 | 0.83 | -0.04 | 0.34 | 0.34 |
| 2 | ↑ | ↓ | ↓ | 0.06 | 0.09 | -0.02 | 0.72 | 0.37 | 0.43 | -0.07 | -0.04 | 0.14 | 0.15 | 0.83 | 0.14 | 0.34 | 0.37 |
| 3 | --- | ↓ | ↓ | 0.07 | 0.09 | -0.02 | 0.8 | 0.37 | 0.44 | -0.08 | -0.06 | 0.14 | 0.15 | 0.81 | 0.01 | 0.34 | 0.34 |
| 4 | ↓ | --- | ↓ | 0.07 | 0.09 | -0.01 | 0.87 | 0.37 | 0.44 | -0.07 | -0.06 | 0.13 | 0.14 | 0.83 | -0.05 | 0.31 | 0.32 |
| 5 | ↑ | --- | ↓ | 0.06 | 0.09 | -0.02 | 0.72 | 0.36 | 0.44 | -0.08 | -0.05 | 0.13 | 0.14 | 0.82 | 0.12 | 0.33 | 0.35 |
| 6 | --- | --- | ↓ | 0.07 | 0.09 | -0.02 | 0.80 | 0.37 | 0.44 | -0.07 | -0.05 | 0.13 | 0.14 | 0.83 | 0.04 | 0.31 | 0.32 |
| 7 | ↓ | ↓ | --- | 0.09 | 0.09 | 0.01 | 1.09 | 0.44 | 0.44 | 0.00 | -0.01 | 0.14 | 0.14 | 1.00 | -0.08 | 0.31 | 0.32 |
| 8 | ↑ | ↓ | --- | 0.08 | 0.09 | -0.01 | 0.90 | 0.44 | 0.43 | 0.00 | 0.01 | 0.14 | 0.14 | 1.00 | 0.11 | 0.32 | 0.33 |
| 9 | --- | ↓ | --- | 0.09 | 0.09 | 0.00 | 1.00 | 0.44 | 0.44 | -0.01 | -0.01 | 0.14 | 0.14 | 0.98 | -0.02 | 0.31 | 0.31 |
| 10 | ↓ | --- | --- | 0.09 | 0.09 | 0.01 | 1.09 | 0.44 | 0.44 | 0.00 | -0.01 | 0.13 | 0.13 | 1.00 | -0.08 | 0.29 | 0.30 |
| 11 | ↑ | --- | --- | 0.08 | 0.09 | -0.01 | 0.90 | 0.44 | 0.44 | 0.00 | 0.00 | 0.13 | 0.13 | 0.99 | 0.09 | 0.30 | 0.32 |
| 12 | --- | --- | --- | 0.09 | 0.09 | 0.00 | 1.00 | 0.44 | 0.44 | 0.00 | 0.00 | 0.13 | 0.13 | 1.01 | 0.01 | 0.29 | 0.29 |

SGA = small-for-gestational age; RD = risk difference; RR = risk ratio, ESE = empirical standard error, RMSE = root mean squared error

[a] Bias, empirical standard error, and root mean squared error were calculated using the natural log-transformed risk ratio and reported on that scale.

**Table S6. Results of the analyses for small-for-gestational-age (SGA) live birth across all 12 scenarios estimated via an Aalen-Johansen estimator, treating fetal death as a competing event and non-SGA live birth as a censoring event.**

| | Effect of Treatment on Risk of… | | | Truth | | | | Aalen-Johansen Estimator | | | | | | | | |
|---|---|---|---|---|---|---|---|---|---|---|---|---|---|---|---|---|
| | | | | Risk | | | | Risk | | Risk Difference | | | | Risk Ratio[a] | | | |
| Scenario | Miscarriage | Preterm live birth | SGA | Initiators | Non-Initiators | RD | RR | Initiators | Non-Initiators | Estimate | Bias | ESE | RMSE | Estimate | Bias | ESE | RMSE |
| 1 | ↓ | ↓ | ↓ | 0.07 | 0.09 | -0.01 | 0.87 | 0.40 | 0.42 | -0.02 | 0.00 | 0.18 | 0.18 | 0.94 | 0.08 | 0.42 | 0.43 |
| 2 | ↑ | ↓ | ↓ | 0.06 | 0.09 | -0.02 | 0.72 | 0.33 | 0.42 | -0.09 | -0.07 | 0.16 | 0.18 | 0.77 | 0.06 | 0.42 | 0.42 |
| 3 | --- | ↓ | ↓ | 0.07 | 0.09 | -0.02 | 0.8 | 0.36 | 0.43 | -0.07 | -0.05 | 0.18 | 0.18 | 0.83 | 0.03 | 0.42 | 0.42 |
| 4 | ↓ | --- | ↓ | 0.07 | 0.09 | -0.01 | 0.87 | 0.40 | 0.42 | -0.02 | -0.01 | 0.18 | 0.18 | 0.94 | 0.08 | 0.41 | 0.42 |
| 5 | ↑ | --- | ↓ | 0.06 | 0.09 | -0.02 | 0.72 | 0.33 | 0.43 | -0.1 | -0.07 | 0.17 | 0.19 | 0.76 | 0.04 | 0.44 | 0.44 |
| 6 | --- | --- | ↓ | 0.07 | 0.09 | -0.02 | 0.80 | 0.37 | 0.42 | -0.05 | -0.04 | 0.18 | 0.18 | 0.86 | 0.07 | 0.42 | 0.42 |
| 7 | ↓ | ↓ | --- | 0.09 | 0.09 | 0.01 | 1.09 | 0.46 | 0.42 | 0.04 | 0.04 | 0.18 | 0.18 | 1.10 | 0.02 | 0.38 | 0.38 |
| 8 | ↑ | ↓ | --- | 0.08 | 0.09 | -0.01 | 0.90 | 0.38 | 0.42 | -0.04 | -0.03 | 0.16 | 0.16 | 0.90 | 0.00 | 0.38 | 0.38 |
| 9 | --- | ↓ | --- | 0.09 | 0.09 | 0.00 | 1.00 | 0.42 | 0.43 | -0.01 | -0.01 | 0.17 | 0.17 | 0.97 | -0.03 | 0.39 | 0.39 |
| 10 | ↓ | --- | --- | 0.09 | 0.09 | 0.01 | 1.09 | 0.46 | 0.42 | 0.04 | 0.03 | 0.17 | 0.18 | 1.10 | 0.01 | 0.37 | 0.37 |
| 11 | ↑ | --- | --- | 0.08 | 0.09 | -0.01 | 0.90 | 0.38 | 0.43 | -0.05 | -0.04 | 0.17 | 0.17 | 0.89 | -0.02 | 0.40 | 0.40 |
| 12 | --- | --- | --- | 0.09 | 0.09 | 0.00 | 1.00 | 0.42 | 0.42 | 0.00 | 0.00 | 0.17 | 0.17 | 1.00 | 0.00 | 0.38 | 0.38 |

SGA = small-for-gestational age; RD = risk difference; RR = risk ratio, ESE = empirical standard error, RMSE = root mean squared error

[a] Bias, empirical standard error, and root mean squared error were calculated using the natural log-transformed risk ratio and reported on that scale.

*Table S7. Results of the analyses for small-for-gestational-age (SGA) live birth across all 12 scenarios estimated via an Aalen-Johansen estimator, treating both fetal death and non-SGA live birth as competing events (modeled separately).*

| | Effect of Treatment on Risk of… | | | Truth | | | | Aalen-Johansen Estimator | | | | | | | | |
|---|---|---|---|---|---|---|---|---|---|---|---|---|---|---|---|---|
| | | | | Risk | | | | Risk | | Risk Difference | | | | Risk Ratio[a] | | | |
| Scenario | Miscarriage | Preterm live birth | SGA | Initiators | Non-Initiators | RD | RR | Initiators | Non-Initiators | Estimate | Bias | ESE | RMSE | Estimate | Bias | ESE | RMSE |
| 1 | ↓ | ↓ | ↓ | 0.07 | 0.09 | -0.01 | 0.87 | 0.07 | 0.09 | -0.01 | 0.00 | 0.01 | 0.01 | 0.87 | 0.00 | 0.11 | 0.11 |
| 2 | ↑ | ↓ | ↓ | 0.06 | 0.09 | -0.02 | 0.72 | 0.06 | 0.09 | -0.02 | 0.00 | 0.01 | 0.01 | 0.72 | 0.00 | 0.11 | 0.11 |
| 3 | --- | ↓ | ↓ | 0.07 | 0.09 | -0.02 | 0.8 | 0.07 | 0.09 | -0.02 | 0.00 | 0.01 | 0.01 | 0.80 | 0.00 | 0.11 | 0.11 |
| 4 | ↓ | --- | ↓ | 0.07 | 0.09 | -0.01 | 0.87 | 0.07 | 0.09 | -0.01 | 0.00 | 0.01 | 0.01 | 0.87 | 0.00 | 0.11 | 0.11 |
| 5 | ↑ | --- | ↓ | 0.06 | 0.09 | -0.02 | 0.72 | 0.06 | 0.09 | -0.02 | 0.00 | 0.01 | 0.01 | 0.72 | 0.00 | 0.11 | 0.11 |
| 6 | --- | --- | ↓ | 0.07 | 0.09 | -0.02 | 0.80 | 0.07 | 0.09 | -0.02 | 0.00 | 0.01 | 0.01 | 0.80 | 0.00 | 0.11 | 0.11 |
| 7 | ↓ | ↓ | --- | 0.09 | 0.09 | 0.01 | 1.09 | 0.09 | 0.09 | 0.01 | 0.00 | 0.01 | 0.01 | 1.09 | 0.00 | 0.10 | 0.10 |
| 8 | ↑ | ↓ | --- | 0.08 | 0.09 | -0.01 | 0.90 | 0.08 | 0.09 | -0.01 | 0.00 | 0.01 | 0.01 | 0.90 | 0.00 | 0.11 | 0.11 |
| 9 | --- | ↓ | --- | 0.09 | 0.09 | 0.00 | 1.00 | 0.09 | 0.09 | 0.00 | 0.00 | 0.01 | 0.01 | 1.00 | 0.00 | 0.10 | 0.10 |
| 10 | ↓ | --- | --- | 0.09 | 0.09 | 0.01 | 1.09 | 0.09 | 0.09 | 0.01 | 0.00 | 0.01 | 0.01 | 1.09 | 0.00 | 0.10 | 0.10 |
| 11 | ↑ | --- | --- | 0.08 | 0.09 | -0.01 | 0.90 | 0.08 | 0.09 | -0.01 | 0.00 | 0.01 | 0.01 | 0.90 | 0.00 | 0.11 | 0.11 |
| 12 | --- | --- | --- | 0.09 | 0.09 | 0.00 | 1.00 | 0.09 | 0.09 | 0.00 | 0.00 | 0.01 | 0.01 | 1.00 | 0.00 | 0.10 | 0.10 |

SGA = small-for-gestational age; RD = risk difference; RR = risk ratio, ESE = empirical standard error, RMSE = root mean squared error

[a] Bias, empirical standard error, and root mean squared error were calculated using the natural log-transformed risk ratio and reported on that scale.

**Methods**

*Methods S1. Formulae for the estimates, bias, empirical standard error, and root mean square error.*

We used formulae as provided by Morris et al. 2019 to calculate the estimates, bias, empirical standard error, and root mean square error for risks and risk differences across 2,000 cohorts for each set of data generating parameters.[30] These same formulae and their statistical definitions are provided below, as written by Morris et al. 2019, for completeness.

| Performance Measure | Definition | Estimate |
|---|---|---|
| Estimate | $E[\hat{\theta}]$ | $\frac{1}{n_{sim}} \sum_{i=1}^{n_{sim}} \hat{\theta}_i$ |
| Bias | $E[\hat{\theta}] - \theta$ | $\frac{1}{n_{sim}} \sum_{i=1}^{n_{sim}} \hat{\theta}_i - \theta$ |
| Empirical Standard Error | $\sqrt{Var(\hat{\theta})}$ | $\sqrt{\frac{1}{n_{sim}-1} \sum_{i=1}^{n_{sim}} (\hat{\theta}_i - \bar{\theta})^2}$ |
| Root Mean Squared Error | $\sqrt{E[(\hat{\theta}-\theta)^2]}$ | $\sqrt{\frac{1}{n_{sim}} \sum_{i=1}^{n_{sim}} (\hat{\theta}_i - \theta)^2}$ |